\documentclass[conference]{IEEEtran}
\usepackage{amssymb}
\usepackage{rotating}

\usepackage{pdflscape}
\usepackage{afterpage}
\usepackage{capt-of}
\usepackage{lipsum}

\usepackage{enumitem} 

\usepackage[table]{xcolor}

\usepackage{placeins}

\usepackage{longtable}

\usepackage{helvet}

\makeatletter
\def\ps@IEEEtitlepagestyle{%
	\def\@oddfoot{\mycopyrightnotice}%
	\def\@evenfoot{}%
}
\def\mycopyrightnotice{%
	{\footnotesize \textbf{Preprint: IEEE/ACM ASONAM 2016, August 18-21, 2016, San Francisco, CA, USA 978-1-5090-2846-7/16/\$31.00 © 2016 IEEE}\hfill}
	\gdef\mycopyrightnotice{}
}

\makeatletter
\let\old@ps@headings\ps@headings
\let\old@ps@IEEEtitlepagestyle\ps@IEEEtitlepagestyle
\def\confheader#1{%
	\def\ps@headings{%
		\old@ps@headings%
		\def\@oddhead{\strut\hfill#1\hfill\strut}%
		\def\@evenhead{\strut\hfill#1\hfill\strut}%
	}%
	\def\ps@IEEEtitlepagestyle{%
		\old@ps@IEEEtitlepagestyle%
		\def\@oddhead{\strut\hfill#1\hfill\strut}%
		\def\@evenhead{\strut\hfill#1\hfill\strut}%
	}%
	\ps@headings%
}
\makeatother

\ifCLASSINFOpdf
   \usepackage{latexsym, amsmath, amsthm}
   \usepackage{algpseudocode}
   \usepackage[ruled,linesnumbered,vlined,norelsize]{algorithm2e} 
   \usepackage{array}
   \usepackage[table]{xcolor}
   \usepackage[hypcap]{caption}
   
   \theoremstyle{definition} 
   \newtheorem{definition}{Definition}
   \newcolumntype{P}[1]{>{\centering\arraybackslash}p{#1}}
     
   \DeclareGraphicsExtensions{.pdf,.jpeg,.png,.eps}
\else
\fi
\begin{document}
%
\title{Investigative Simulation: Towards Utilizing Graph Pattern Matching for Investigative Search}

\thispagestyle{plain}
\pagestyle{plain}

\author{\IEEEauthorblockN{Benjamin W. K. Hung, Anura P. Jayasumana }
\IEEEauthorblockA{Department of Electrical and Computer Engineering\\
Colorado State University\\
Fort Collins, Colorado, USA\\
Email: \{benjamin.hung, anura.jayasumana\} @colostate.edu}}


%


\maketitle

\begin{abstract}
This paper proposes the use of graph pattern matching for \textit{investigative graph search}, which is the process of searching for and prioritizing persons of interest who may exhibit part or all of a pattern of suspicious behaviors or connections. While there are a variety of applications, our principal motivation is to aid law enforcement in the detection of homegrown violent extremists. We introduce \textit{investigative simulation}, which consists of several necessary extensions to the existing dual simulation graph pattern matching scheme in order to make it appropriate for intelligence analysts and law enforcement officials.  Specifically, we impose a categorical label structure on nodes consistent with the nature of indicators in investigations, as well as prune or complete search results to ensure sensibility and usefulness of partial matches to analysts. Lastly, we introduce a natural top-\textit{k} ranking scheme that can help analysts prioritize investigative efforts. We demonstrate performance of investigative simulation on a real-world large dataset.

\end{abstract}


%
\IEEEpeerreviewmaketitle

\section{Introduction} \label{introduction}
In the era of big data and user-generated content, one of the most pressing needs continues to be filtering irrelevant data and finding the desired information to make timely and accurate decisions \cite{Ma2015}. Since much of the data in a variety of domains can be conveniently represented as heterogeneous data graphs, graph pattern matching is of growing importance to find such information. While a recent overview \cite{Ma2015} lists complex object identification, software plagiarism detection, and traffic route planning as some additional applications, the bulk of the research in this field is oriented towards social search and recommender systems \cite{Basu-Roy2013} \cite{Choudhury2013} \cite{Fan2010} \cite{Fan2013} \cite{Khan2013} \cite{Ma2014} \cite{Pienta2014} \cite{Tong2007}. In social search, for instance, one may utilize graph pattern matching to find an entity with specific types of connections or attributes, while recommender systems help individuals form collaboration networks with people with specific skills and expertise. 

\textbf{Investigative Graph Search}. This paper proposes the use of graph pattern matching for \textit{investigative graph search} and introduces several necessary extensions to existing graph pattern matching schemes in order to make them appropriate for intelligence analysts and law enforcement officials. We consider \textit{investigative graph search} (or simply \textit{investigative search}) as a special case of social search, and define it as a process of searching for and prioritizing persons of interest who may exhibit part or all of a pattern of suspicious behaviors or connections. Some distinguishing characteristics of investigative graph search from other searches include: 

\begin{enumerate} [leftmargin=*]
	\item Nodes in a query pattern are hypothesized indicators of a latent behavior of interest; not all indicators may appear in the matched result to make a partial match worthy of further investigation.
	\item Some indicators nodes are only significant in the context or presence of other indicators.
	\item The ranked full or partial match results should help analysts prioritize among potentially many matches based upon the presence of red-flag indicators as well as the similarity of the matches to the query.
\end{enumerate}

\begin{figure}[!ht]
	\centering
	\includegraphics[width=3.5in]{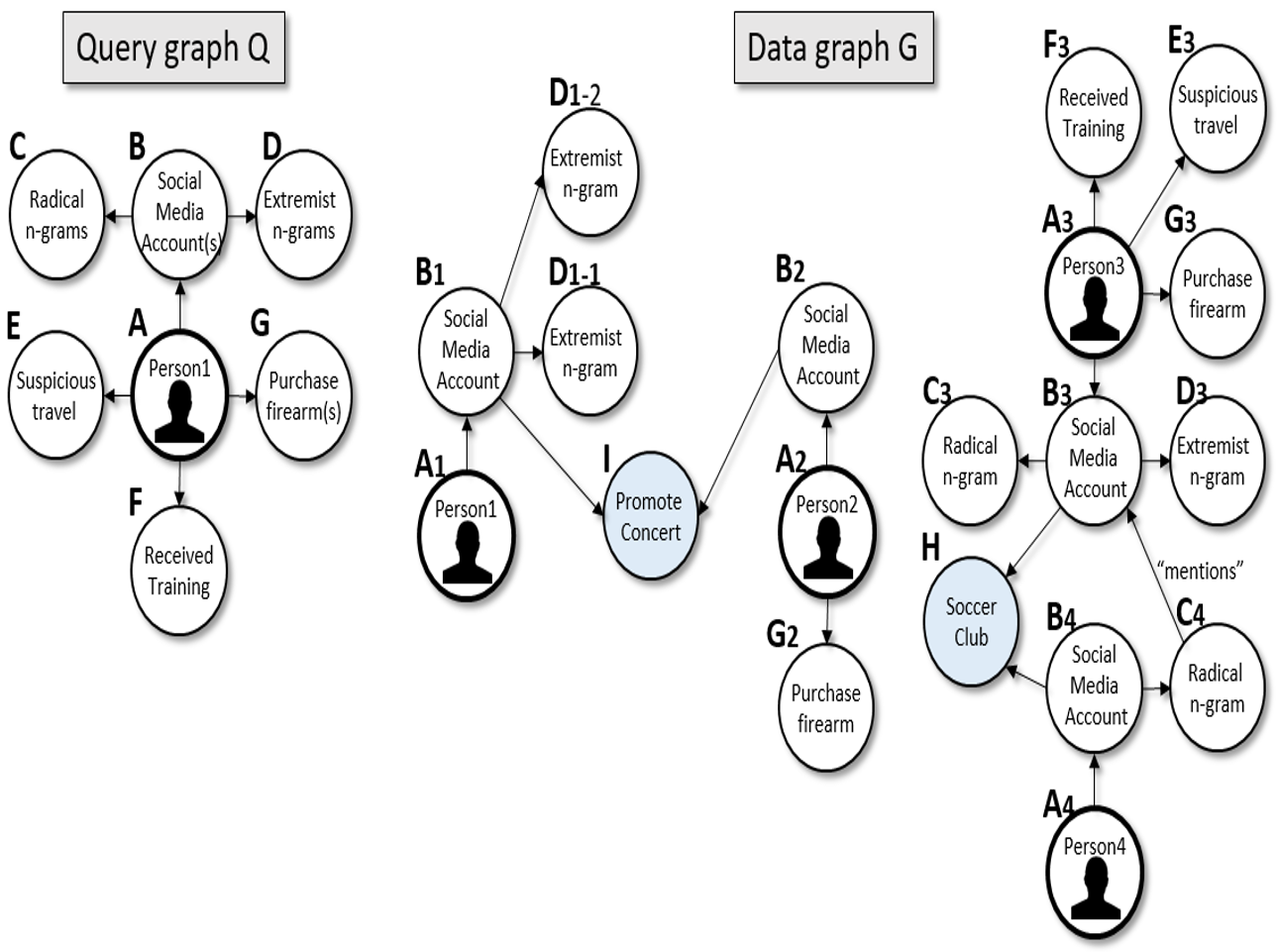}\\
	(a) \hspace{2in}(b) \\
	\centering
	\includegraphics[width=2.75in]{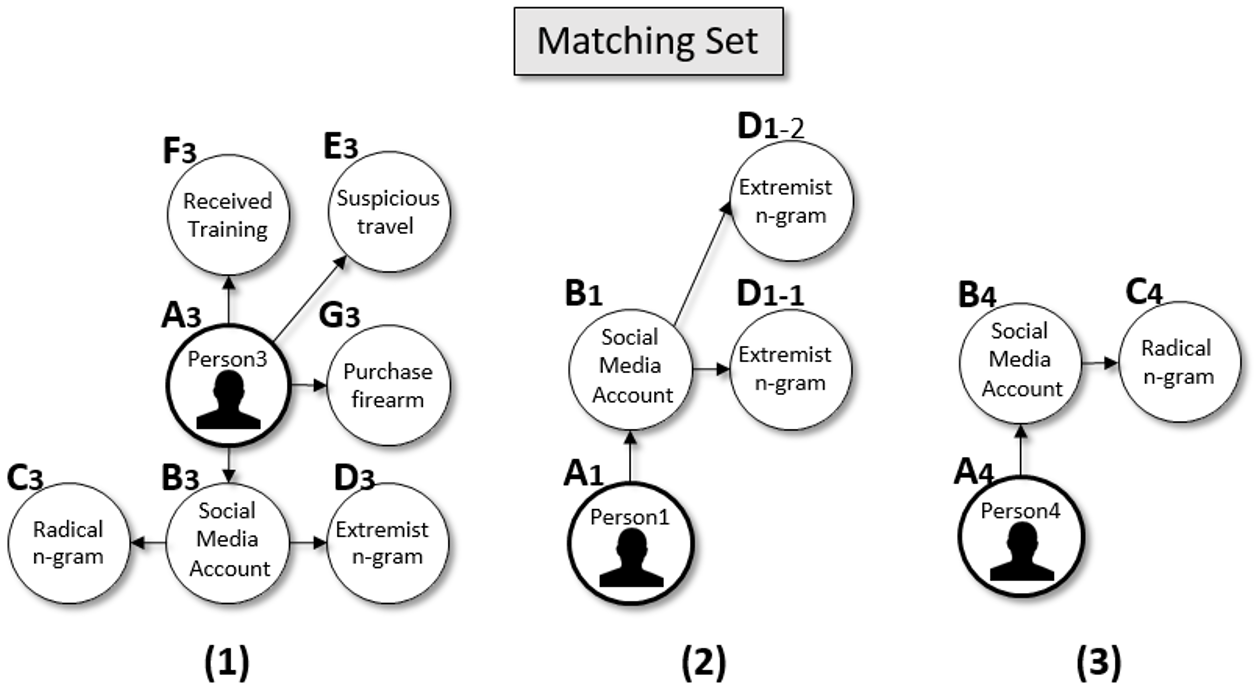}\\
	(c)
	\caption{An example graph query of a potential homegrown violent extremist (a) and a fictitious data graph of 4 people with on and off-line activities (b). Nodes in the data graph represent distinct entities (e.g., person or social media account) or behaviors (e.g., posting extremist $n$-gram, purchasing a firearm, etc.) with the class label shown inside the node. The letter of the label outside a node is a code for the class, and the number (if applicable) denotes the person responsible with that entity or behavior. The query graph represents the pattern of nodes by class that may help identify potential homegrown violent extremists. Fig. 1(c): Desired matching set via investigative simulation that includes full and partial matches in rank order.}
	\label{MotivatingExampleGraphandQuery}
\end{figure}

While investigative graph search can be used in a variety of applications from the detection of signs for post-traumatic stress in veterans to tracking customer journey indicators, our primary motivation is to aid in the detection of homegrown violent extremists (HVEs) who seek to commit acts of terrorism in the United States and abroad. 

Facilitated and inspired by extremist organizations utilizing the internet and social media for recruitment and radicalization, HVEs have presented significant challenges to law enforcement in recent years. In 2011 the National Counterterrorism Center (NCTC) published the results of a study that examined the behavioral indicators of 27 recent cases \cite{NCTC2011} and identified a total of 16 specific behaviors that were present in at least half the cases. This included not only active online participation in blogs or chat rooms, but also communication with extremists, consumption of jihadist videos and propaganda, suspicious foreign travel or attempted travel, and expressions of acceptance or intent to conduct violent jihad or martyrdom operations. Despite the identification of these behavioral indicators, the risk of threats going undetected to law enforcement remains high. The FBI Director recently requested help from local law enforcement to keep track of HVEs, saying ``It's an extraordinarily difficult challenge task to find -- that's the first challenge -- and then assess those who may be on a journey from talking to doing'' \cite{PerezandProkupecz2015}.   

\textbf{A Motivating Example}. A small example problem related to the investigative search for homegrown violent extremists is shown in Fig. \ref{MotivatingExampleGraphandQuery}. A simplified query graph $Q$ of some possible indicators of a homegrown violent extremist is shown in (a). The pattern is a person who 1) posted radical- and extremist- labeled $n$-grams from a social media account, 2) underwent suspicious travel to a foreign country and received terrorist-related training, and 3) purchased a firearm. A simplified data graph $G$ of 4 people each with various on- and off-line activities is shown in (b).\footnote{While not a focus of this paper, it is important to note that our approach is predicated on access to and proper classification/labeling of data from open and restricted sources to produced large heterogeneous data graphs. For example, we envision that social media data (e.g., Twitter, Facebook) would be fused with firearm background check databases and local/state/federal criminal and terrorist databases (including data from the FBI's Tripwire and `FBI Tips' programs as well as the TSA's Automated Targeting System and Secure Flight programs).} The problem is to find all whole or partial matches of the query $Q$ in the data graph $G$ and present results according to some intuitive ranking scheme.

Dual simulation \cite{Ma2014}, which represents the state of the art in graph pattern matching through simulation-based approaches short of the match locality constraint, technically returns only Person 3 (and his related activities) as the only matching connected subgraph. Because the algorithmic implementation \cite{Ma2014} of dual simulation also returns remnant individual node matches, the \textit{n}-grams from Person 1 and 4 ($D_{1-1}, D_{1-2}, C_4$) as well as the `purchase firearm' node ($G_2$) from Person 2  are partial matches. From this, we identify three shortcomings of dual simulation for investigative search: 1) the requirement for every node in the query to have some match, and no allowance for partial matches, is too restrictive when an investigator may include indicator nodes in the query which need not be associated with every person of interest, 2) any remnant node matches with valid matching indicators do not contain the subject of the search, and 3) remnant node matches may contain nodes that are innocuous activities except when observed with other suspicious indicators.    

Our modification to dual simulation, which we call \textit{investigative simulation}, is designed to address these aforementioned shortcomings, and returns the match result in Fig. \ref{MotivatingExampleGraphandQuery}(c). Person 3 (and his related activities) is still the only complete match for all indicators. However, Persons 1 and 4 (and their related activities) are now also returned as partial matches due to the posting of radical and extremist $n$-grams. Note that none of Person 2's activities ('purchase firearm' or social media account nodes) are now returned because neither indicator is important unless there are other suspicious indicators of motivation or intent to commit targeted violence. 

In order to return such a match result from the dual simulation pattern matching schemes, we determined the need to impose a categorical structure on nodes consistent with the nature of indicators in the threat assessment literature, as well as the need to prune or complete search results to ensure sensibility and minimize false positives. Lastly, we introduce an intuitive ranking scheme to help investigators prioritize results based upon the presence of red-flags. 

\textbf{Our Contributions.} Our two main contributions are:
\begin{enumerate} [leftmargin=*]
	\item We propose \textit{investigative simulation}, an extension of the dual simulation graph pattern matching method that is specialized for investigative search, and propose an algorithm to produce more sensible matches of potential subjects for further investigation.
	\item We propose a roadmap of further advancements necessary to investigative search, particularly for detecting the radicalization of homegrown violent extremists in open source on- and off-line behavior.
\end{enumerate}

The rest of this paper is organized as follows: Section II provides an overview of related work in this field. Section III reviews definitions and notation. Section IV defines the investigative simulation method of graph pattern matching. Section V presents the results of the application of our algorithm on a real world dataset. Section VI provides two case studies related to homegrown violent extremists and the modeling context for our investigative search approach on heterogeneous networks. Finally, Section VII concludes the paper and outlines some recommendations for future work in investigative simulation.    

\section{Related Work} \label{relatedwork}
Our work builds upon advances in graph pattern matching in the static setting. Several surveys exist, including \cite{BhargaviandSupreehi2012} and \cite{Gallagher2006}. Of the two principal types of matching, exact and inexact, we focus our efforts on the state of the art in inexact matching due its flexibility for returning results in the presence of noise or errors in the data \cite{Gallagher2006}. The notable works in static inexact matching include `best-effort matching' \cite{Tong2007}, TALE \cite{Tian2008}, SIGMA \cite{Mongiovi2010}, NeMa \cite{Khan2013}, and MAGE \cite{Pienta2014}. The `inexact' component of these works primarily involves the allowance for finding nearby matches for nodes in which an exact match does not exist. 

Of these, the work most closely related to ours in intention is \cite{Pienta2014}, which first introduced a graph pattern matching method that supports exact and inexact queries on both node and edge attributes as well as wildcard matches. This matching method specifically cites intelligence analysis as a use-case and offers great flexibility in the query construction, allowing analysts to explore the unknown or uncertain connections. However, this matching scheme still does not truly support uncertain indicator-type matches or innocuous nodes that become significant only in the context of other indicators. 

Equally important are simulation-based matching schemes, starting with bounded graph simulation \cite{Fan2010} \cite{Fan2012}, to find meaningful matches given a pattern graph with arbitrary or specified path lengths in the connections. Later, dual and strong simulations \cite{Fan2012} \cite{Ma2014} were developed to preserve query graph topology through enforcement of both parent and child relationships in the match and the imposition of locality constraints. 

For the purposes of investigative search where a person may exhibit an indicator behavior one or more times, we find that simulation-based approaches may be most appropriate due to the allowance of each query node to be matched to multiple nodes in the data graph as long as match labels are preserved at the match-level, as well as with the parent- and child-levels. However, as previously mentioned in Section \ref{introduction}, dual simulation has several shortcomings with investigative search, including the lack of allowance for partial matches, incomplete remnant node matches, and lack of ability to handle matches of innocuous activities that become important only when observed with other suspicious indicators.

Lastly, because most graph queries end up returning many matches given a large graph, researchers have also devised ways to rank the most relevant matches using various goodness functions. Such criteria include social impact \cite{Fan2013}, social diversity \cite{Fan2013}, structural similarity \cite{Basu-Roy2013} \cite{Khan2013} \cite{Tong2007}, weighted attribute similarity \cite{Basu-Roy2013}, and label similarity \cite{Khan2013}. As sophisticated as these ranking methods are, we find that none account for intuitive red-flag indicators (i.e., those matches which demand the immediate attention of an analyst) that are relevant in investigative searches. 

\section{Technical Preliminaries and Notation} \label{techprelim}

Before we define investigative simulation in the next section, we first review graph terminology and notation, as well as the dual simulation graph pattern matching method.

\begin{definition}\textbf{Graph} \cite{Ma2014}.
	In our work, both the query graph $Q$ and data graph $G$ are identically defined as a directed graphs of the form $G(V,E,L)$, where $V$ is a set of nodes, $E\subseteq V \times V$ is a set of edges, in which $(u,u')$ denotes an edge from node $u$ to $u'$; and $L: V \cup E \rightarrow \Sigma$ is a labeling function which assigns nodes and edges to a set of labels $\Sigma$.
\end{definition}

\begin{definition}\textbf{Dual Simulation} \cite{Ma2014}
Graph $G$ matches a pattern $Q$ via dual simulation if there exists a binary match relation $S_D \subseteq V_Q \times V_G$ such that: 
	\begin{itemize} [noitemsep,topsep=0pt,leftmargin=*]
		\item for all nodes $u \in V_Q$ there exists a node $v \in V_G$ such that $(u,v) \in S$; and
		\item for each pair $(u,v) \in S, u \sim v$ (i.e., $L_Q(u)=L_G(v)$), and for each edge $(u, u') \in E_Q$ there exists an edge $(v, v') \in E_G$ such that $(u', v') \in S_D$, and for each edge $(u', u) \in E_Q$ there exists an edge $(v', v) \in E_G$ such that $(u', v') \in S_D$.
	\end{itemize}
	We then refer to $S_D$ as a match (via dual simulation) to $Q$.
\end{definition}

Dual Simulation was a significant advancement over graph simulation and previous notions because it preserved not only parent-child relationships, but also child-parent relationships in the match and thus produced more sensible matches. However, as described in Sections \ref{introduction} and \ref{relatedwork}, there are shortcomings when performing investigative search. We summarize the notations we utilize in this paper in Table \ref{NotationTable}.  

\begin{table}[!h]
	\begin{center}
		\caption{Summary of notations}
		\label{NotationTable} 
		\begin{tabular}{|P{1in}|p{2in}|}
			\hline
			\rowcolor{gray!50} \textbf{Notation}&\textbf{Description/Meaning}\\\hline
			$G$&Data graph $G(V_G,E_G,L_G)$\\\hline
			$Q$&Query graph $Q(V_Q, E_Q, L_Q)$\\\hline
			$S_D$&Binary Match relation (via dual simulation)\\\hline
			$S_{\textit{InvSim}}$&Match relation (via investigative simulation)\\\hline
			$(u,u')$& Directed edge from node $u$ to $u'$\\\hline
			$u\in V_Q$, $v \in V_G$ & Nodes with index $u$ ($v$) are in graph $V_Q$ ($V_G$), respectively.\\\hline
			$R_{(u,v)}$& Relevant set of matching node $v \in V_G$ w.r.t. query node $u \in V_Q$.\\\hline														
		\end{tabular}
	\end{center}
\end{table}

\begin{table*}[t]
	\centering
	\setlength\extrarowheight{2pt}
	\begin{tabular}{|p{3.2cm}|p{4.5cm}|p{1.1cm}|p{4.7cm}|p{2.5cm}|}
		\hline
		\rowcolor{gray!50} \textbf{Node type (Abbreviation)} & \textbf{Definition} & \textbf{Source(s)} & \textbf{Example(s)} & \textbf{Sufficient for further investigation} \\
		\hline
		\textbf{Query focus (QF)} 
		& Subjects of an investigative query (i.e., we want individuals who match a particular pattern).
		& \cite{Fan2013}
		& 1) Person
		& No \\	
		\hline
		\textbf{Individually innocuous but related activity (IIRA)} 
		& Behavior that is individually not harmful or threatening (and often common), but in this context may be part of threat when combined with something else.
		& N/A
		& 1) Buying a gun \newline 2) Having a social media account
		& No \\
		\hline
		
		\textbf{Indicator (IND)} 
		& Term used to broadly classify those unusual behaviors which may suggest that an individual is be a threat and work more as potential building blocks towards a threat assessment. Can encompass behaviors, traits, characteristics, risk factors, and warning signs. 
		& \cite{Depue2007} \cite{SchuurmanandEijkman2015}
		& 1) Fascination with weapons \newline
		2) Purchasing large quantities of fertilizer \newline
		3) Downloading the Anarchist Cookbook \newline
		4) Re-tweeting a violent jihadist video \newline
		5) Watching a video from a violent extremist preacher
		& No \\
		\hline
		
		\textbf{Red flag indicator (RF)}
		& Risk factors which, if present, will singly determine that a case ranks as a high risk or concern until proven otherwise \cite{Meloy2011}.
		& \cite{Meloy2011} \cite{White2014}
		& 1) Motives for violence \footnotemark \newline
		2) Homicidal ideas \newline
		3) Fantasies or preoccupations \newline
		4) Violent intentions or expressed threats \newline
		5) Pre-attack planning and preparation \newline
		6) Received terrorist training overseas
		& Yes \\
		\hline
	\end{tabular}
	\caption{Categorical node labels for investigations}
	\label{CategoryNodeTable}
\end{table*}

\section{Investigative Simulation}

The proposed investigative simulation approach is described next. We first describe an important categorical node labeling method specific to investigative queries, provide a formal definition of investigative simulation, and then propose an algorithm for this new matching technique.  

\subsection{Categorical node labeling for investigations}
We found that existing dual simulation technique over-matches in investigative search because it returns matching nodes that were not properly qualified in the query. Specifically, investigative queries may contain nodes that are representative of perfectly legal and innocuous activities that are only potential indicators of a latent behavior of interest when they occur with other indicators. For example, purchasing a firearm may only serve as a targeted violence threat indicator if it is accompanied by an overt communication of threats to others. At the same time, it is possible to identify indicators, should they occur, which may \textit{be individually sufficient} to warrant further investigation. While we could use a numerical node weighting scheme to ensure such indicator/node differences (as in \cite{Basu-Roy2013} and \cite{Yang2014}), we suspect that node weights may change when in the context of other indicators and as such complicate the matching process. For example, while we may initially want the indicator for purchasing a firearm to have a relatively low weight when it is the only indicator of a person as a threat, such an indicator would likely be weighted much more heavily when it also occurred with the same person making overt threats \cite{Depue2007}. 

\begin{figure}[!htbp]
	\centering
	\includegraphics[width=1.50in]{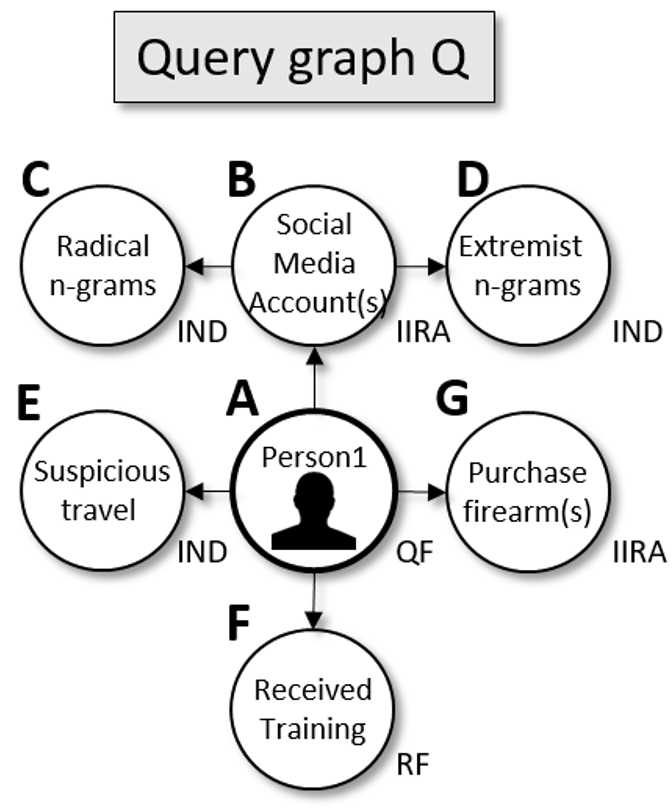}
	\caption{Consistent with the node category definitions and examples, we label the nodes in the example graph query pattern $Q$ as follows: query focus (A), individually innocuous but related activity (B,G), indicator (C, D, and E), and red flag indicator (F).}
	\label{MotivatingExampleQueryWithCat}	
\end{figure}

\footnotetext{See \cite{White2014} for the definitions of example indicators 1-5.}

We thus propose a categorical weighting of nodes to the set of node/edge labels $\Sigma$ based upon research from the threat assessment and homegrown violent extremist radicalization literature, which has generally advocated for condition-based weighting. See Table \ref{CategoryNodeTable} for the definitions and examples provided. The first category \textit{query focus (QF)} is used to label nodes which are the subject of the investigative query-- namely the people in the data graph. The second category \textit{individually innocuous but related activity (IIRA)} is for activities which need to occur in conjunction with other more suspicious indicators to be worth further examination. The third category \textit{indicator (IND)} is a broad term to classify unusual behaviors which may suggest a person is a threat. Lastly, the fourth category \textit{red flag indicator (RF)} is for activities which are \textit{individually sufficient} to warrant further investigation. If a node does not fall in any of these categories, it can be labeled as \textit{no category (NC)}. This initial modeling extension allows us to address the non-sensible matches from dual simulation through algorithmic modifications, which we describe in the following section.

\subsection{Definition of Investigative Simulation}

\begin{definition}\textbf{Investigative simulation: An extension of dual simulation for investigative search}
	
	Graph $G$ contains partial or complete matches of pattern $Q$ if there exists a binary match relation $S_{\textit{InvSim}} \subseteq V_Q \times V_G$ such that: 
	\begin{itemize} [noitemsep,topsep=0pt,leftmargin=*]
		\item for all nodes $u \in V_Q:L(u) = \textit{`QF'}$ and at least 1 node $u \in V_Q:L(u) = \textit{`IND'}$ or $\textit{`RF'}$ there exists a node $v \in V_G$ such that $(u,v) \in S_{\textit{InvSim}}$;
		\item for each pair $(u,v) \in S_{\textit{InvSim}},$ where $u \in V_Q$ and $v \in V_G$, $u \sim v$ (i.e., $L_Q(u)=L_G(v)$), and 
		\item for each edge $(u, u') \in E_Q$ there exists an edge $(v,v') \in E_G$ such that $(u',v') \in S_{\textit{InvSim}}$, and for each edge $(u',u) \in E_Q$ there exists an edge $(v',v) \in E_G$ such that $(u',v') \in S_{\textit{InvSim}}$.
	\end{itemize}
	We then refer to $S_{\textit{InvSim}}$ as a match (via investigative simulation) to $Q$.
\end{definition}

Instead of all nodes in $Q$ needing a match in $G$ (as in dual simulation), investigative simulation allows for partial matches by only requiring all `QF' nodes and at least 1 indicator node (regular `IND' or red-flag `RF') to have a match in $G$. This keeps matching results specific to `QF' nodes with at least a single indicator-type that may make it worthy of further analysis or investigation.  

\subsection{InvSim- Extension of dual simulation algorithm} 

One of the merits of the `DualSim' algorithm for dual simulation found in \cite{Ma2014} was that it returns the entire binary match relation $S_D$, which contains not only complete matches (connected component subgraphs for all nodes in $Q$) but also remnant node matches (those nodes whose parent and/or child were pruned away as a result of their connections). While \cite{Ma2014} never uses these remnant node matches in the construction of the maximum subgraph, they in fact form the basis of the partial matches that are informative for investigative searches. However, they are by nature incomplete matches (e.g., in the case of the network schema in the motivating example problem in Fig. \ref{MotivatingExampleGraphandQuery}, the query focus nodes associated with remnant indicator nodes were not in the match relation). We develop a post-processing extension to the dual simulation algorithm (Algorithm 1: InvSim, short for \underline{Inv}estigative \underline{Sim}ulation) that corrects both issues specific to indicator-type patterns. In it we utilize a modified 2-hop concept of a relevant set from \cite{Fan2013}. Given a match $v$ of a query node $u$ in $V_Q$, the relevant set of $v$ w.r.t. $u$ (denoted as $R_{(u,v)}$) includes all matches $v'$ of $u'$ for up to the 2-hop descendants $u'$ of $u$ in $V_Q$.  

\begin{algorithm}[!b]
	\KwIn{Query graph $Q$ with investigation category node labels, and data graph $G$} 
	\KwOut{The match relation $S_{\textit{InvSim}}$ of $Q$ and $G$}
	\BlankLine
	\ForEach{$u \in V_Q$}{
		$sim(u):=$\textbraceleft $v$ \textbar $v \in V_G$ \textbf{and} $L_Q(u) = L_G(v)$\textbraceright
	}
	\While{\upshape there are changes}{
		\ForEach{\upshape edge $(u,u') \in E_Q$ and \textbf{each} node $v \in sim(u)$}{
			\lIf{\upshape there is no edge $(v,v')$ in $G$ with $v'\in sim(u')$}{
				$sim(u):=sim(u)$\textbackslash \textbraceleft $v$\textbraceright
			}
		}
		\ForEach{\upshape edge $(u',u) \in E_Q$ and \textbf{each} node $v \in sim(u)$}{
			\lIf{\upshape there is no edge $(v',v)$ in $G$ with $v'\in sim(u')$}{
				$sim(u):=sim(u)$\textbackslash \textbraceleft $v$\textbraceright
			}
		}
		\lIf{$sim(u)= \emptyset$}{\textbf{return} $\emptyset$}
	}
	$S_D:=$ \textbraceleft $(u,v)$\textbar $u\in V_Q, v\in sim(u)$\textbraceright\\
	\ForEach{\upshape node $v\in S_D$ where $L(v)=$`QF'}{
		\If{\upshape $L(\tilde{v})=$`IIRA' for all $\tilde{v} \in R_{(u,v)} \cap S_D$}{
			$sim(\tilde{u}):=sim(\tilde{u})$\textbackslash \textbraceleft $\tilde{v}$\textbraceright;
			\lIf{\upshape node $v\in S_D$}{
				$sim(u):=sim(u)$\textbackslash \textbraceleft $v$\textbraceright				
			}
		}
		\If{\upshape there exists a node $\tilde{v} \in R_{(u,v)} \cap S_D$ \textbf{and} $v \notin S_D$}{
			$sim(u):=sim(u)$ $\cap$ \textbraceleft $v$\textbraceright;
			\lIf{\upshape every node $\check{v}$ along the shortest path from $(v,\tilde{v})$ is not in $S_D$}{
				$sim(\check{u}):=sim(\check{u})$ $\cap$ \textbraceleft $\check{v}$\textbraceright, $\forall (\check{u},\check{v})$
			}
		}		  
	}
	$S_{\textit{InvSim}}:=$ \textbraceleft $(u,v)$\textbar $u\in V_Q, v\in sim(u)$\textbraceright\\
	\textbf{return} $S_{\textit{InvSim}}$.    
	\caption{InvSim (for \underline{Inv}estigative \underline{Sim}ulation)}
\end{algorithm}

Lines 1-9 are those lines found in \cite{Ma2014}, which is the implementation of the dual simulation algorithm and results in the intermediate match relation $S_D$. Our post-processing extension to this algorithm begins in Line 10, when we iterate through all nodes in the matching set in data graph $G$ which are labeled as `QF' or query focus (i.e. persons). If the intersection of the relevant set of $v$ and the match relation $S_D$ (nodes $\tilde{v}$) are all of type `IIRA' (individually innocuous but related activity), then we remove this node from the match relation and remove its query focus parent if it was in the relation (Lines 11-12). This effectively removes matching nodes that are considered benign without the presence of other indicators, as well as the associated person from further consideration. 

Next, we ensure that the parent query focus nodes of other matching nodes of type `indicator' are included in the match relation $S_D$. We first search for all nodes $\tilde{v}$ that are both in the relevant set of $v$ and the match relation $S_D$ but whose parent query focus node $v$ is not yet in $S_D$ (Line 13). We join this node $v$ to the match relation (Line 14) as well as add all nodes in the shortest path from node $v$ to $\tilde{v}$ in the match relation (Line 14). Finally, we consolidate and return the modified match relation $S_{\textit{InvSim}}$ (Line 15-16).

\section{Results}

\subsection{Real dataset for a proxy investigative search}
\begin{figure}[!ht]
	\centering
	\raisebox{-0.5\height}{\includegraphics[width=1.7in]{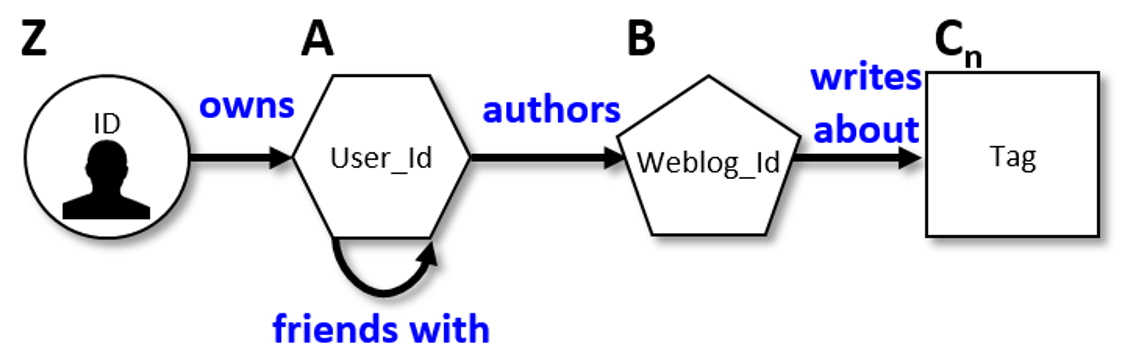}}
	\raisebox{-0.5\height}{\includegraphics[width=1.74in]{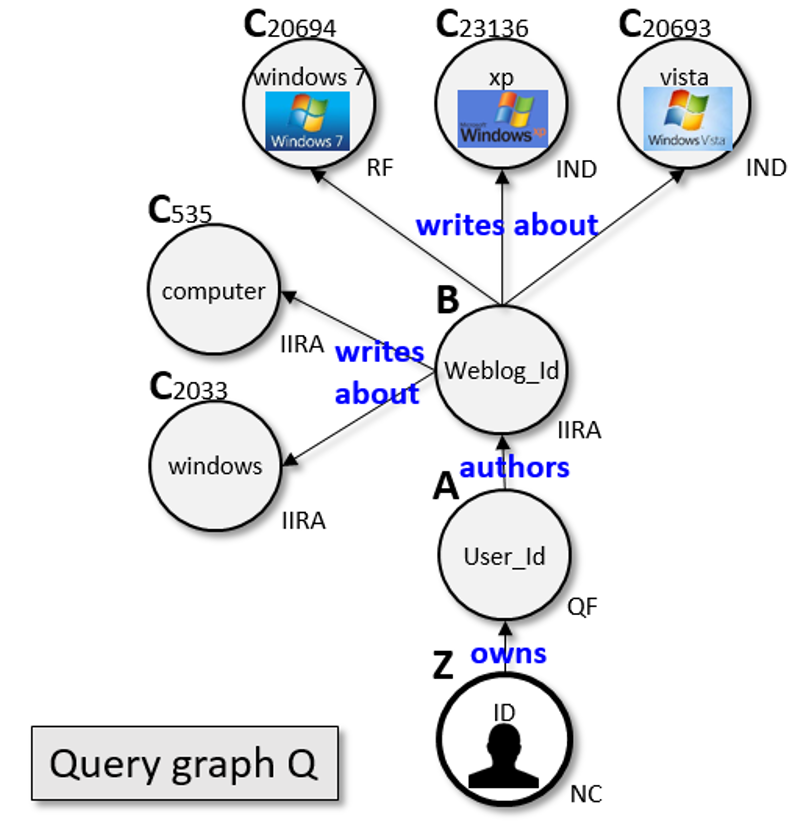}}
	(a) \hspace{1.5in} (b)
	\caption{Network schema of the BlogCatalog graph (a). IDs own account User\_Ids. User\_Ids author one or more Weblog\_Ids, and are friends with other User\_Ids. Weblog\_Ids write about one or more tags (which are user specified). Experimental query for BlogCatalog data (b). Query focus is for User\_Ids who had been writing blogs broadly related to `computers' and `windows', and specifically to Windows operating systems. In this example, we treat the tag `windows 7' as a red flag indicator. }
	\label{BlogCatalogSchema}	
\end{figure}

In order to test investigative simulation on real data, we utilized the BlogCatalog dataset,\footnote{Available at http://dmml.asu.edu/users/xufei/datasets.html} which is a scrape taken in July 2009 of a social media site that allows users to register and promote their own blog and connect with other bloggers. The graph had over 470,000 nodes and over 4 million edges; it is further detailed in Table \ref{BlogCatalogGraphCharacteristics}. The network schema shown in Fig. \ref{BlogCatalogSchema}(a) describes the node types and connections present the network. In essence, an ID owns a User\_Id, which in turn both authors blogs with a Weblog\_Id as well as forms directed friendship connections with other User\_Ids. Lastly, each weblog will provide one or more user-specified tags.  

\begin{table}[!ht]
	\begin{center}
		\caption{BlogCatalog Graph Characteristics}
		\label{BlogCatalogGraphCharacteristics} 
		\begin{tabular}{|p{2in}|P{1in}|}
			\hline
			\rowcolor{gray!50} \textbf{Characteristics}&\textbf{Value}\\\hline
			\rowcolor{gray!20} \textbf{Total Nodes}&\textbf{471,267}\\\hline
			\hspace{.5em} Number of ids&88,781\\\hline
			\hspace{.5em} Number of userids&80,949\\\hline
			\hspace{.5em} Number of weblogs&127,227\\\hline
			\hspace{.5em} Number of unique tags&174,310\\\hline
			\rowcolor{gray!20} \textbf{Total Edges}&\textbf{4,098,290}\\\hline
			\hspace{.5em} Number of links from id to userid&88,784\\\hline
			\hspace{.5em} Number of links from userid to userid&3,223,640\\\hline
			\hspace{.5em} Number of links from userid to weblog&127,227\\\hline
			\hspace{.5em} Number of links form weblog to tags&658,639\\\hline							
		\end{tabular}
	\end{center}
\end{table}

\subsection{Query Description} To test the performance of the matching scheme and algorithm, we devised a proxy query on a benign subject matter with structural parallels to investigations as shown in Fig. \ref{BlogCatalogSchema}(b). The query focus is for user IDs who had been writing blogs related to Microsoft Windows operating systems (XP and/or Vista) and subsequently also began to write about Windows 7 when it was released in July 2009 (month in which the data was collected). Node Z is a true person ID, node A is the user ID query focus, and node B is the weblog with certain tags. All C nodes are meant to be seen as labels of a post or blog entry (i.e., determined through machine-classified semantic analysis). The labels `computer' (C535) and `windows' (C2033) are IIRA (i.e., relatively frequent labels which help provide context or additional clarity on the true topic set), and labels `xp' (C23136) and `vista' (C20693) are indicators that the blog is about Windows operating systems (i.e., necessary but not sufficient for trajectory behavior). Finally, label `windows 7' (C20684) is considered a red flag indicator. 


\subsection{Ranking Method and Analysis}
As expected, investigative simulation returned meaningful partial matches to the query. Our intuitive ranking scheme for the top-\textit{k} results was to 1) first order by the presence of any `QF' nodes with red-flag (`RF') indicators, and 2) followed by the size of the relevant matching set for each `QF' node (i.e., in decreasing order of \textbar $R_{(u,v)}\cap S_{\textit{InvSim}}$\textbar, where $L_Q(u)=L_G(v)=\textit{`QF'}$). This method effectively highlights to analysts those first who have red-flag indicators, followed by the those who have the most indicators towards the latent behavior of interest. 

\begin{figure}[!ht]
	\centering
	\includegraphics[width=3.3in]{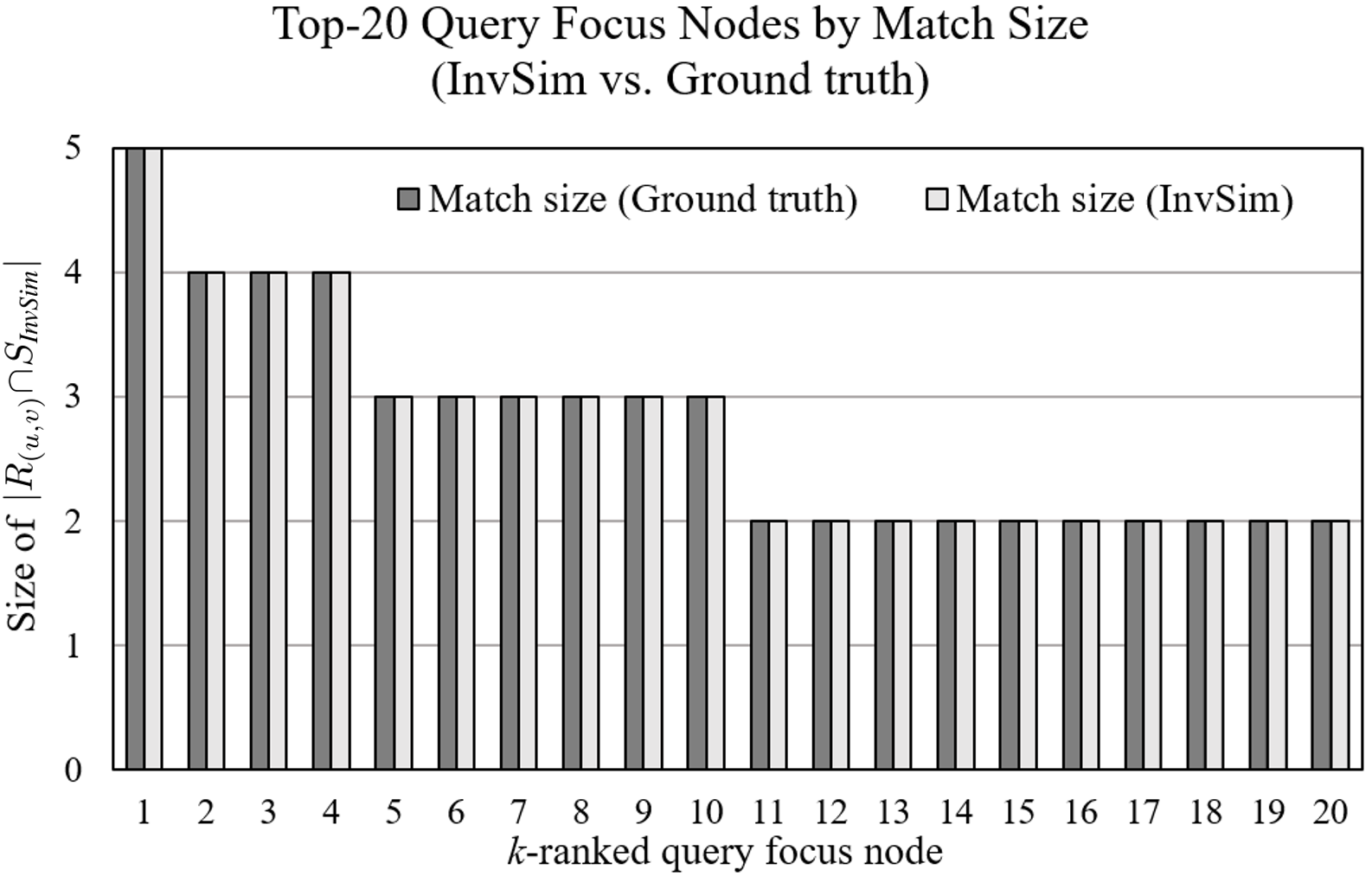}\\
	(a)\\
	\includegraphics[width=3.0in]{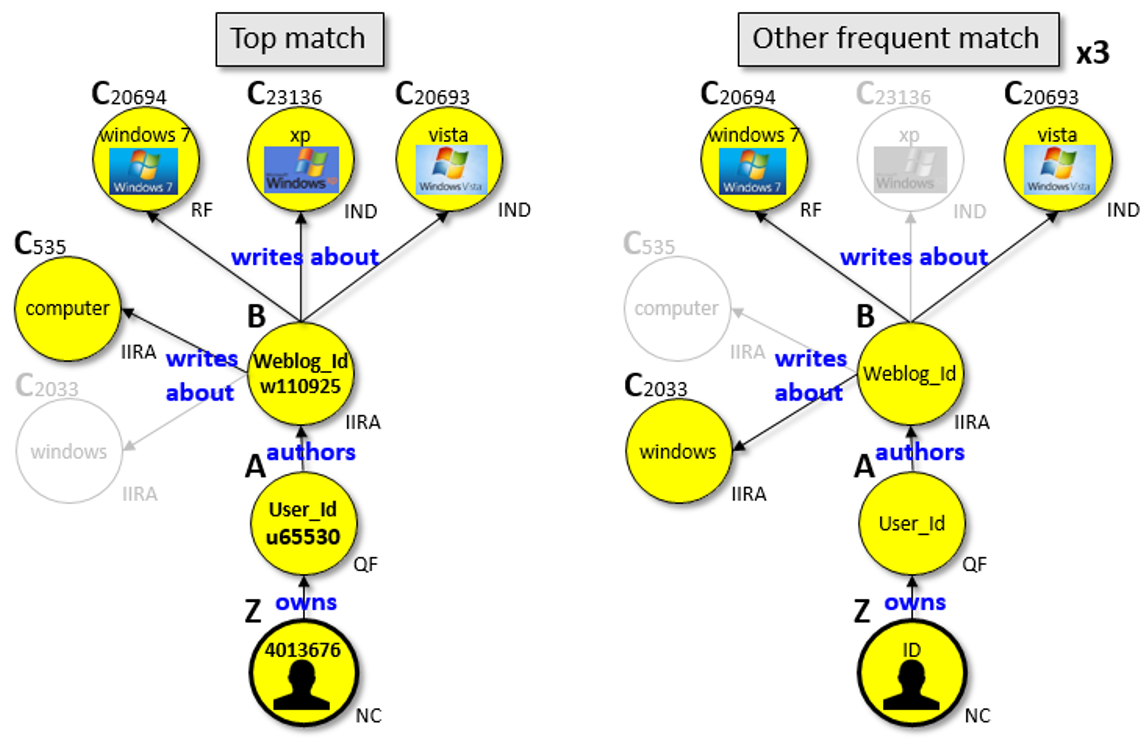}\\
	\hspace{.23in}(b) \hspace{1.32in} (c)\\
	\caption{A paired bar graph (a) showing the exact correspondence of the top-20 query focus nodes by match size between both InvSim and exhaustive search, where \textbar $R_{(u,v)}\cap S_{\textit{InvSim}}$\textbar \hspace{.1em}is the number of matching nodes in the relevant set of each query focus node.  Fig. 4(b-c): Top-4 results of investigative simulation on the BlogCatalog dataset with the query in Fig. 3(b). The top-match (b) is User\_Id `u65530' with 5 indicator nodes matching in the relevant set (2 directed hops from Node A). Note the presence of the red flag indicator `windows 7` in each of these matches. The grayed-out nodes were the original query nodes not matched.}
	\label{BlogCatalogSubgraphResults}
\end{figure}

We find that investigative simulation performed well in the matching, as measured with both quantitative and qualitative methods. First, we quantitatively measured the similarity between the top-20 results with the original query pattern by using the Jaccard similarity, and compared it with the top-20 ground truth results acquired through exhaustive search. The paired bar graph in Fig. \ref{BlogCatalogSubgraphResults}(a) shows the exact correspondence. Qualitatively, we performed subjective validation of the sensibility of each of the top-10 match results. The top-4 partial matches to the query are shown in Fig. \ref{BlogCatalogSubgraphResults}(b-c).


\section{Two Case Studies of Homegrown Violent Extremism} \label{CaseStudies}
In this section, we present two short case studies of recent homegrown violent extremism to provide real-world context to the modeling of on- and off-line behaviors as heterogeneous data graphs, as well as to our investigative search approach. While these graph-based connections were established after the plot or attack in a subsequent law enforcement investigation, we also aspire to demonstrate the analysis possible if there were a better fusion of law enforcement and public security databases with open-source social media.    

\begin{figure}[!ht]
	\centering
	\includegraphics[width=3.56in]{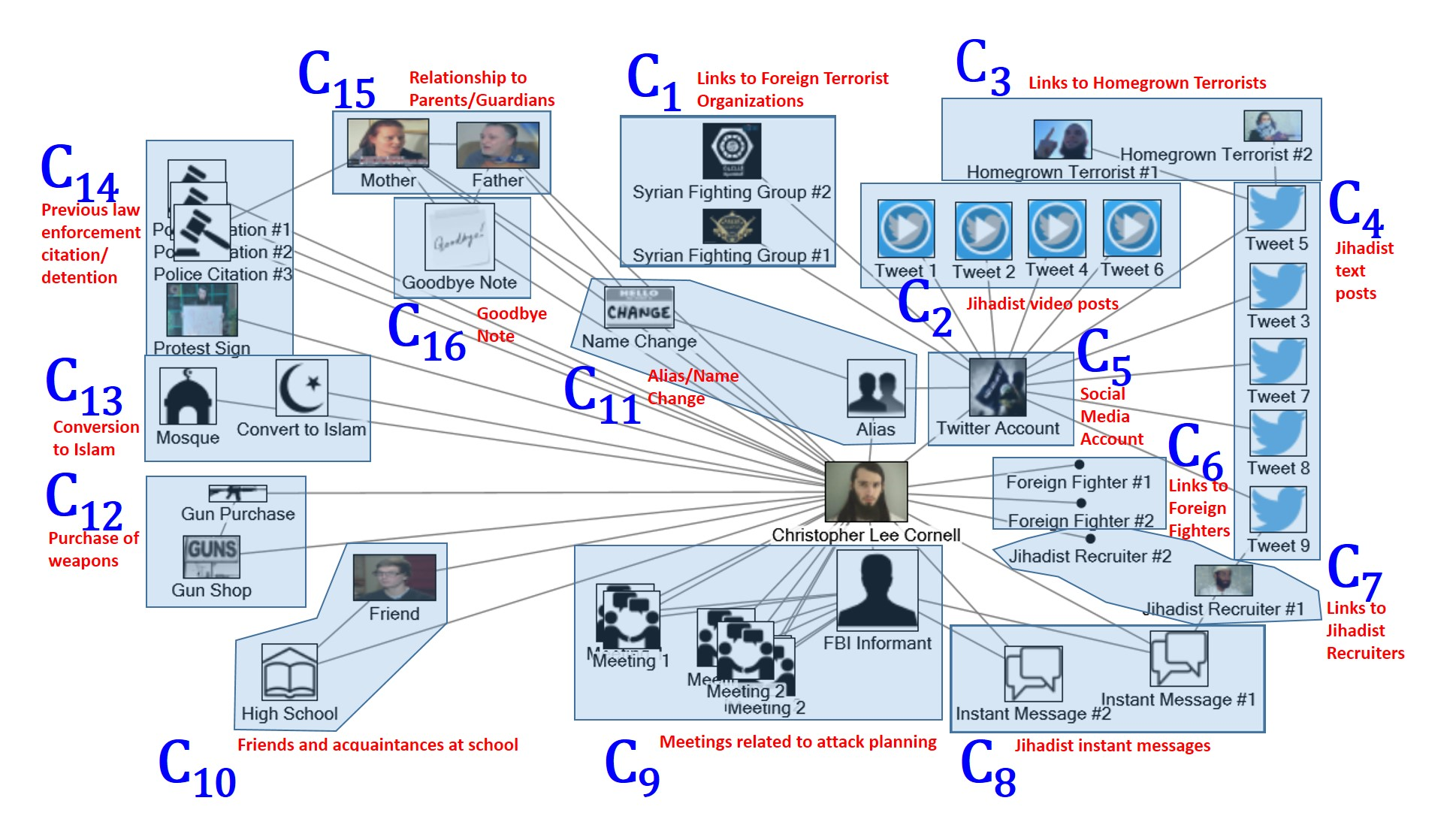}
	\caption{US Capitol Attack Plot, 2015. Example class graph of Christopher Lee Cornell showing the indicators and signals of his radicalization and progress towards an attack.}
	\label{LeeClassGraph}
\end{figure}

\textbf{Case Study 1.} Christopher Lee Cornell, an example of a recent homegrown violent extremist in the United States, was arrested by the FBI in January 2015 for allegedly planning to employ pipe bombs at the U.S. Capitol and then open fire on nearby people. From the criminal complaint \cite{LeeCriminalComplaint2015} and other open sources \cite{Katz2015}, we employed a methodology called process-tracing to identify increasing indicators of radicalization that ultimately led up Lee's purchase of weapons to use in a planned attack. The indicators include Lee's purported activity on Twitter with references to jihadist recruiter Al Awlaki and other terrorists, posting of Islamic State propaganda videos, and his attack planning with an FBI confidential human source. When the signals and indicators are combined into a heterogeneous data graph, there are a total of 43 nodes and 16 discernible classes in the class graph as shown in Fig. \ref{LeeClassGraph}  (produced in NodeXL \cite{Smith2010}).  These classes are devised from the political science literature on potential indicators of homegrown terrorists \cite{Gill2014} \cite{Klausen2015} \cite{Meloy2011} \cite{SchuurmanandEijkman2015} \cite{SilberandBhatt2007}. 


\textbf{Case Study 2.} The next case study is the San Bernardino, California terrorist attack on December 2, 2015. The perpetrators Syed Farook and wife Tashfeen Malik conducted a mass shooting and attempted bombing that killed 14 people and injured 22 at the Inland Regional Center. Enrique Marquez has also been charged with conspiring to provide material support to terrorists \cite{MarquezCriminalComplaint2015}. Just as in the Lee case, critical signals in this case were embedded in the perpetrators' social media posts. For example, nearly a month before the attack, Marquez purported posted this exchange on Facebook with another user: ``No one really knows me. I lead multiple lives and I'm wondering when its all going to collapse on M[e]...Involved in terrorist plots, drugs, antisocial behavior, marriage, might go to prison for fraud, etc.'' \cite{MarquezCriminalComplaint2015}. In future work, we propose to include \textit{n}-grams or key words as nodes in the data graph. Connections to those nodes from the social media posts serve to link the most suspicious phrases and words that may warrant further investigation from law enforcement officials. 

\begin{figure}[!t]
	\centering
	\includegraphics[width=3.5in]{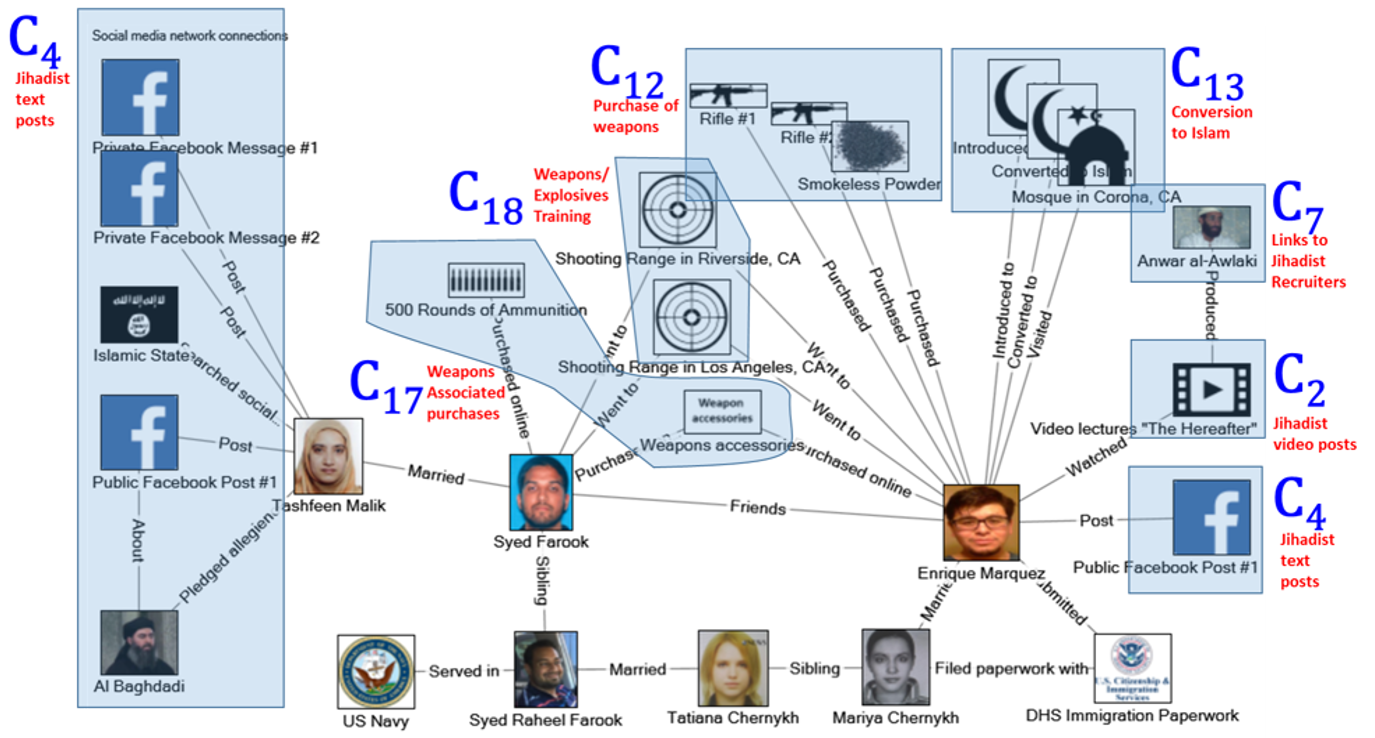}
	\caption{San Bernardino Terrorist Attack, 2015. Example class graph of Syed Farook, Tashfeen Malik, and Enrique Marquez showing the indicators and signals of their collective radicalization and preparations for the attack.}
	\label{SanBernardinoClassGraph}
\end{figure}

It is also interesting to note that the indicators of radicalization and attack preparations were not present in just a single individual, but in all \textit{three} (as shown in Fig. \ref{SanBernardinoClassGraph}). This case study points out the need to explore conspiratorial graph patterns (where there is match complementarity over more than one query focus node), which is not currently addressed by any extant matching techniques. As part of our ongoing work in investigative graph simulation, we propose to identify these types of conspiracy cells through query-focus node cluster matching.    

\textbf{Insights for Real-World Investigative Search}. These two case studies provide a real-world context to the challenges of applying graph pattern matching to aid in the search for homegrown violent extremists. In particular, constructing the query is problematic because almost all existing graph pattern matching approaches rely on certainty in the query and do not have a categorical node labeling structure for indicators. Between the two case studies, one can discern the variability in the presence of indicators as well as the significant number of nodes which might be classified as `individually innocuous but related activities.' Investigative simulation seeks to address these issues by allowing for partial matches of a comprehensive indicator query, suggesting node categories for investigative searches, and pruning or augmenting the match relation to produce sensible matches and fewer false positives. 

\section{Conclusion and Future Work}
The availability of user-generated content in the form of blogs, micro-blog posts, and videos has led many researchers to study the potential for detecting latent signals of human behavior for a variety of business and security-related purposes. In this paper, we have introduced investigative graph search as a process of searching for and prioritizing persons of interest who may exhibit part or all of a pattern of suspicious behaviors or connections. We also propose the technique of investigative simulation as an extension of dual simulation for investigative searches. We show that this form of graph pattern matching produces more sensible matches and more complete partial matches through the imposition of categorical node labels related to indicators, as well as offer an algorithm to find the matches. 

There are many planned areas of future work in investigative search and investigative simulation which we will briefly mention here. First, we intend to find more efficient implementations of the InvSim matching algorithm that use a combination of indexing techniques and early termination approaches to find the top-\textit{k} matches. Second, as identified in the San Bernardino case study in Section \ref{CaseStudies}, we intend to formulate the problem and devise efficient algorithms for match complementarity to help identify conspiratorial plots and threats. 

Beyond investigative search in the static setting, we ultimately seek to incorporate graph dynamics with the imposition of timestamps on edges in both the query graph $Q$ and the data graph $G$ (building upon the work in \cite{Song2014}). This important extension would not only provide analysts with the ability to match by time or sequence, but would also give them a sense of the pace of indicator connections in latent behaviors of interest.

\section*{Acknowledgment}
The lead author wishes to thank the General Omar N. Bradley Foundation for the 2016 Research Fellowship in Mathematics.

\end{document}